# Study of intercalation/deintercalation of $Na_xCoO_2$ single crystals

C.T. Lin[1*], D.P. Chen[1,2], P. Lemmens[1,3], X.N. Zhang[1], A. Maljuk[1], and P.X. Zhang[1,2]

[1] Max-Planck-Institut für Festköperforschung, Heisenbergstr. 1, D-70569 Stuttgart, Germany
[2] Institute of Advanced Materials for Photoelectronics, KUST, Yunnan, 650051, China
[3] Institute for Physics of Condensed Matter, TU Brauschweig, 38106 Braunschweig, Germany

Single crystals of $Na_xCoO_2$ with β-phase (x=0.55, 0.60 and 0.65), α'-phase (x=0.75) and α-phase (x=0.9, 1.0) have been grown by the floating zone technique. The Na-extraction and hydration were carried out for the α'-sample to get superconducting phase of $Na_xCoO_2 \bullet yH_2O$ (x~0.3, y~1.3). Hydrated single crystals exhibit cracked layers perpendicular to the *c*-axis due to a large expansion when the water is inserted into the structure. A study of intercalation/deintercalation was performed to determine the stability of the hydrated phase and effects of hydration on the structure of the compound. X-ray diffraction and Thermogravimetric experiments are used to monitor the process of water molecules accommodated in and removed from the crystal lattice. The initial intercalation process takes place with two-water molecules (corresponding to y=0.6) inserted in a formula unit, followed by a group of four (y=1.3) to form a cluster of $Na(H_2O)_4$. Thermogravimetric analysis suggests that the deintercalation occurs with the removal of the water molecules one by one from the hydrated cluster at elevated temperatures of approximately 50, 100, 200 and 300 °C, respectively. Our investigations reveal that the hydration process is dynamic and that water molecule inter- and deintercalation follow different reaction paths in an irreversible way.

[*]Corresponding author: email, ct.lin@fkf.mpg.de, tel. 0049-711-6891458, fax, 0049-711-6891093

# 1. Introduction

It is of particular interest that the $Na_xCoO_2$ compound can be transferred to a superconductor with water molecules intercalated between $CoO_2$ layers [1-3] as the observation of superconductivity in a transition metal oxide is a rare and exceptional event. A lot of efforts have been made to study the mechanism of superconductivity, magnetic susceptibility, and to investigate basic properties, as specific heat, electronic anisotropy [4-8], but even these are far from being understood. To obtain superconductivity in the compound a controlled doping level or the ratio of $Co^{3+}/Co^{4+}$ has to be achieved and the intercalated water should also be considered as an important parameter. Although neutron and x-ray powder diffraction were used to determine the crystal structure and the arrangement of water molecules in the sodium layer [9-11], it is rather difficult to fix the local coordination of the water molecules. Furthermore, the intercalated water in the crystal is extremely unstable; hence the accurate characterization of the compound becomes another problem to be solved [12-15]. Several models were proposed [10, 11] to interpret the structure of the compound. It was concluded by J.D. Jorgensen et al. that the four $D_2O$ molecules coordinated to a Na-ion between two neighboring $CoO_2$ layers to form the superconducting phase of $Na_{0.3}CoO_2 \bullet yD_2O$ (y=1.3). However, J.W. Lynn et al [10] reported that the water molecules in the compound resemble to the structure of ice. There are several plausible scenarios of unconventional, non-phonon superconductivity in $Na_{0.3}CoO_2 \bullet yH_2O$, and it is expected that the water coordination also affect the electronic properties of the $CoO_2$ planes. Moreover, the hydrated compound is extremely unstable under ambient conditions [5, 12] or at elevated temperatures and exhibits several phases with different **y** values by dehydration. Therefore, the availability of single crystals is important, that allow an accurate characterization of the compound with respect to its physical, chemical, electrical and thermal properties.

In this work, we present a crystal growth method of growing large and high quality single crystals. The x-ray diffraction (XRD) and Raman scattering experiments are used to reveal the change of the crystal structure during the intercalation process by immersing the sample into water. The details of the water deintercalation from the crystal lattice are demonstrated by Thermogravimetric (TG) on heating the fully hydrated sample.

**2. Experiment**

Single crystals were grown in an optical floating zone furnace (Crystal System Incorporation, Japan) with 4 X 300 W halogen lamps installed as infrared radiation sources. Starting feed and seed materials were prepared from $Na_2CO_3$ and $Co_3O_4$ of 99.9% purity. The feed rod consisted of a Na and Co mixture with a nominal composition of $Na_xCoO_2$, where x=0.55, 0.60, 0.75, 0.90 and 1.0 respectively. Well-mixed powders of the required composition were loaded into alumina crucible and heated at 750 °C for a day. The heated powders were reground and calcined at 850 °C for another day. They were then shaped into cylindrical bars of ~φ6x100 mm by pressing at an isostatic pressure of ~70 MPa and then sintered at 850 °C for one day in flowing oxygen to form feed rods. The sintered feed rod was premelted at the velocity of 27 mm/h by traveling the upper and lower shafts, respectively. This procedure results in a pre-densification of the feed rod. After premelting the ~20mm long rod was cut and used as a first seed and hereafter the grown crystal was used as a seed. The feed rod and the growing crystal were rotated at 15 rpm in opposite directions. In an attempt to reduce the volatilization of Na and obtain stoichiometric and large crystals, we applied traveling rates of 2 mm/h under pure oxygen flow of 200 ml/min throughout the growing procedure.

To carry out the Na-extraction process, the specimens were cut from the crystal ingot with composition of $Na_{0.75}CoO_2$ and placed in the oxidizing agent $Br_2/CH_3CN$ for around 100 hours, and then rinsed out in acetonitrile. The change of sodium content of the resulting crystals is generally proportional to the bromine concentration in the $CH_3CN$ agent [12]. Crystals of $Na_xCoO_2$ with x=0.3 result from the 6.6 mol $Br_2/CH_3CN$ treatment.

The XRD measurements were carried out with the x-ray diffractometer (Philips PW 1710) using Cu $K_\alpha$ radiation to analyze the phase purity as well as to determine the crystal structure and the lattice parameters. A scanning rate of 0.02 degrees per second was employed to perform the $\theta$ - $2\theta$ scans between 5 and 120 or 5 and 20 degrees, upon experiment requirements. The lattice parameters obtained from XRD data were refined using the commercial program *PowderCell*. The intercalation (hydration) experiment was done by immersing $Na_{0.3}CoO_2$ crystals into de-ionized (D.I.) water and allowed the *001* surface well exposed to air and scanned by x-ray. For the deintercalation a fully hydrated single crystal was loaded in a Pt crucible and then heated in the DTA-TG apparatus (NETZSCH STA-449C) at a slow heating rate of 0.3 °C/min up to 800 °C in flowing oxygen. The analysis of DTA-TG was performed to estimate the water loss from a full hydrate to dehydrate sample.

Raman scattering experiments were performed using a triple spectrometer (DILOR XY) and a liquid nitrogen-cooled CCD. The incident light was provided by an argon ion laser with $\lambda=514.5$ nm. Single crystals of $Na_xCoO_2\bullet yH_2O$ where freshly cleaved within the *001* surface and immediately cooled down in a cryostat to prevent any loss of water from the surface or decomposition due to a local change of Na content. This procedure does not allow a continuous monitoring of the hydration process. Therefore, we have concentrated on comparing the Raman spectra of samples with the hydrated ($y\sim1.3$) and the overhydrated phase ($y\sim1.8$). In previous Raman scattering studies the effect of hydration and sample degradation has been studied [4, 14].

**3. Results and discussion**

*3.1 Crystal growth*

Large single crystals of $Na_xCoO_2$ were obtained with variant Na content under a growth rate of ~2 mm/h in flowing oxygen atmosphere. During growth, we observed that the molten zone was stable and easy to form the $\alpha$'-phase $Na_xCoO_2$ with sodium content of x=0.75 and $\alpha$-phase with x=0.90 and 1.0, while it was hard to grow the $\beta$-phase with x=0.55, 0.60 and 0.65 respectively. The analysis of EDX indicates that as-grown $Na_xCoO_2$ with lower sodium content of x<0.65 consists of multi phases like, $Na_2O$, $Co_3O_4$ and Na-poor phases. Figure 1(a) shows an as-grown typical crystal ingot of $Na_{0.75}CoO_2$. The XRD *00l* diffraction patterns indicate a nearly pure phase of both $\alpha$'- and $\alpha$-$Na_xCoO_2$ crystals, as shown in figure 1(b). Both crystals are identified to be hexagonal structure with two sheets (prismatic, P2) and trigonal with three sheets

(octahedral, O3) of edge-shared $CoO_6$ octahedra in a unit cell for the α'-phase and α-

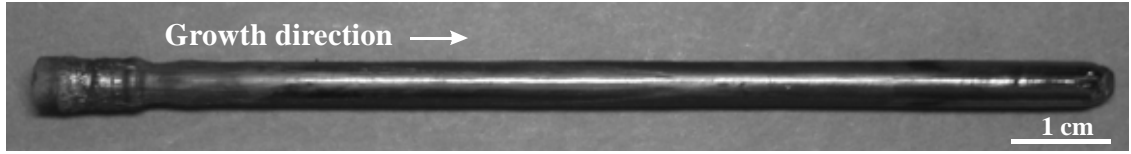

(a)

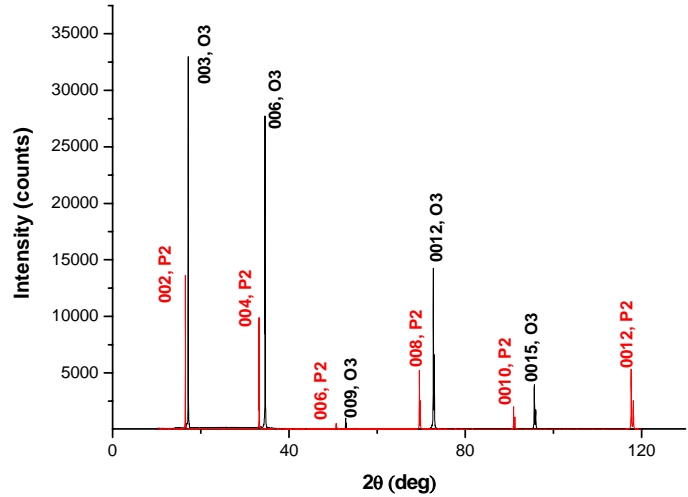

(b)

FIG. 1 (a) A typical α'-phase $Na_{0.75}CoO_2$ single crystal ingot obtained by optical floating zone technique. (b) The *001* XRD patterns showing the pure α'-phase (P2) $Na_{0.75}CoO_2$ and α-phase (O3) $NaCoO_2$, respectively. The split peaks are $CuK_{α1}$ and $CuK_{α2}$ for the higher and lower intensity, respectively.

phase, respectively. The lattice parameters and cell volume were determined by least squares refinement of XRD data and correspond to the space group $P6_3/mmc$ with the lattice constant $a=2.830(2)$ Å, $c=10.98(3)$ Å for the $Na_{0.75}CoO_2$ (P2) and R3/mh with $a=2.899(3)$ Å, $c=15.56(3)$ Å for the $NaCoO_2$ (O3), respectively. The composition distribution along the ingot of the $Na_{0.75}CoO_2$ was determined using EDX. The crystal was scanned through a segment of 3 mm along the growth direction for the determination of Na/Co compositions. The average value of the Na composition was calculated with four measured points in the central and peripheral region of the crystal. It was found that the Na content varied with the temperature fluctuations during growth. At the beginning of the growth the temperature fluctuations were high and thus caused a high variation of the composition, $\Delta x \sim 0.11$, determined for the part 2 cm apart from the seeding part of the ingot. After the seeding part is completed the variation of Na content is small, with $\Delta x \sim 0.06$ determined throughout the ingot. The volatilized white $Na_2O$ powder was seen to cover the inner wall of the silica tube when the growth lasted longer. In general, the Na content gradually decreases with the growing time and the volatilization of $Na_2O$ is observed throughout the entire growth. The white powder could be also observed on the surface of crystals when stored under ambient conditions after growth. This is due to the decomposition of the compound surface reacted with water in air. Therefore the grown crystals must be stored in an evacuated container or a desiccator to avoid decomposition.

*3.2 Na-extraction*

By chemically extracting additional sodium from the structure of $Na_{0.75}CoO_2$ followed by hydration, we obtain the superconducting phase of the compound with the composition $Na_xCoO_2 \bullet yH_2O$ ($0.26 \leq x \leq 0.42$, $y=1.3$). The details are described in Ref. [13]. The crystal of $Na_xCoO_2$ with $x=0.3$ could result from the 6.6 mol $Br_2/CH_3CN$ but a long extraction time, over a week, is needed. Comparing to 1 to 5 days extracting time for the powders [1, 3] the crystal needs longer time to complete the extraction. This can be attributed to that the Na-extraction process takes place along the *100* direction, and the chemical bonding period of Na-O-Na in the NaO layer for single crystal is an order of magnitude longer and more perfect than powders of which are approximately in nanometer scale and randomly distributed. Before and after extracting treatment the sodium composition distribution across the crystals was determined by EDX. Figure 2 is the plot of the sodium distribution in the central area along the crystal growth direction. This analysis indicates that the change of sodium content of the resulting crystals after deintercalation was $\Delta x \sim 0.3$. The sodium intercalant layer expanded with decreasing Na content, because the Na removal results in Co oxidation ($Co^{3+}$ ions oxidized to smaller $Co^{4+}$ ions) and thus the $CoO_2$ layers are expected to shrink [12]. This suggests that a decreased bonding

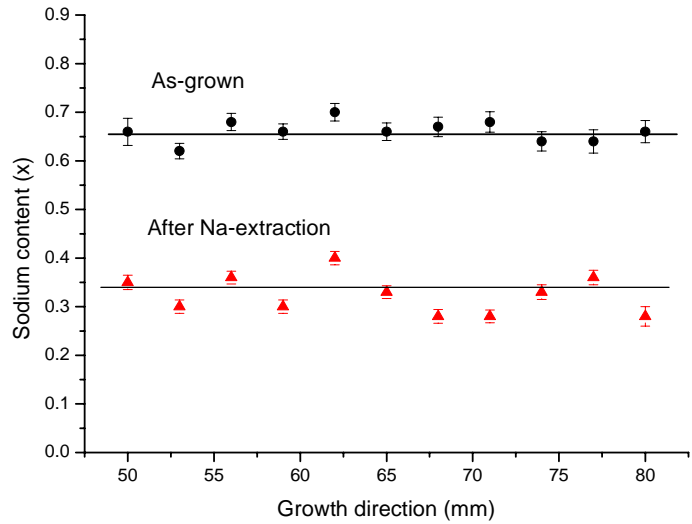

FIG. 2 Sodium distributions along the growth direction for the as-grown and after Na-extraction of the $Na_{0.75}CoO_2$ single crystal.

interaction between layers with decreasing Na content may result in a readily cleaving plane.

*3.3 Crystal structure*

The Na-extracted samples are then hydrated by immersing into D.I. water at room temperature to obtain superconducting phase. After hydration a large increase in thickness is visible to the naked eyes and the morphology exhibits layered cracks perpendicular to the *c* axis. Figures 3 (a) and (b) illustrate the layered structures of the non- and fully hydrated phases of the compound, respectively. According to the layered structure of $Na_xCoO_2$ [11], 24 Co-O bonds are estimated in the (*001*) plane of the $CoO_2$ layer in a unit cell and are of high bonding energy and therefore the layer is robust structurally, while the bonding energy is much weaker in the NaO layer because the $Na^+$ mobility is high and the number of bond Na1-O is 6 and Na2-O only 4/3. Therefore the Na layer readily collapses for the $CoO_2$ layer to terminate as the outer-most surface of the crystal. Moreover, the insertion of water induces two additional intercalated layers of $H_2O$ between NaO and $CoO_2$, i.e., from Co-Na-Co to Co-$H_2O$-Na-$H_2O$-Na-Co, which results in the expansion of the *c* lattice constant from 11.15 to 19.63 Å. The hydrogen in the new intercalated $H_2O$ layer bonds extremely weakly to the NaO and CoO layers. Therefore this leads to the compound being exceptionally unstable with the change of thermal, mechanical or humidity environment. Whatever, typical crystal morphologies are shown in figures 4 (a) and (b), forming by Na-extraction and $H_2O$-intercalation, respectively.

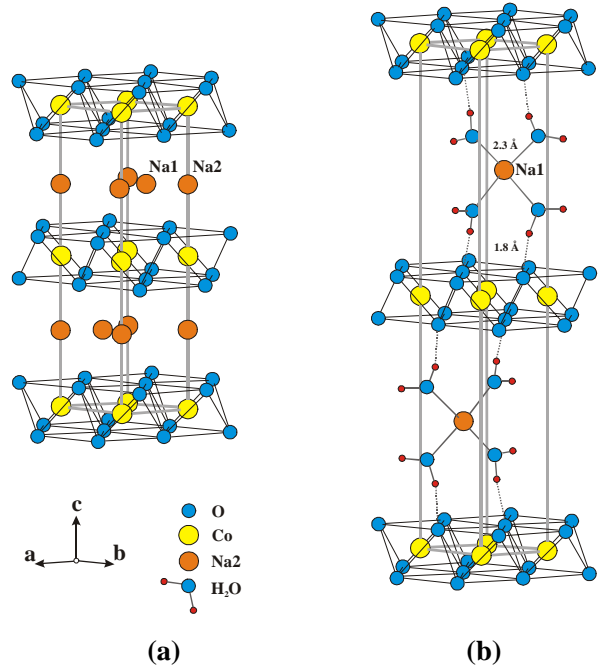

FIG. 3 Schematic drawing of the layered structures [11] for (a) $Na_{0.61}CoO_2$ and (b) water intercalated $Na_{0.33}CoO_2 \cdot 1.3H_2O$.

*3.4 Possible hydrate phases*

The compound of $Na_xCoO_2$ is a layered structure. The layers of $CoO_2$ are formed by octahedral $CoO_6$ with strong bonding and are structurally robust. The compound has two partially occupied sodium (Na1, Na2) sites in the same plane sandwiched between layers of edge-sharing $CoO_6$ octahedra. The ability to change the Na content has been studied [1, 11, 13, 17] and revealed that $Na_{0.3}CoO_2$ can be achieved by the Na extraction starting from $Na_{0.75}CoO_2$. The intercalation of water molecules can then be carried out in $Na_{0.3}CoO_2$ to form $Na_{0.3}CoO_2 \cdot 1.3H_2O$, which gives Na-O bond nearly equal to those in the parent compound. Thus the compound becomes superconducting. According to the structural configuration of an inserted H$_2$O molecule in fig. 3(b), one hydrogen is bonded to an oxygen atom in the CoO$_2$, the oxygen atom and second hydrogen lied in a plane between the Na and Co layers. Such intercalation of water may largely separate the CoO$_2$ layers and expands the c-axis parameters. Many work [3, 11, 12, 13, 17] show that the possible ordered phases of Na$_{0.3}$CoO$_2$•yH$_2$O with the inserted water content of y=0, 0.3, 0.6, 0.9, or 1.3, corresponding to the water molecule number n=0, 1, 2, 3, or 4 per formula unit. Table 1 lists a group of 5 hydrates including y=1.8 (n=6). The XRD patterns of figure 6 show the mixture of hydrates with y=0, 0.3, 0.6, 0.9 and 1.8, when the sample is treated by a hydration time of over 15 days, whereas y=1.3 is missing. Our experiments demonstrate that a further increasing hydration time up to 42 days all hydrated phases mostly collapse or the compound is nearly all decomposed into Na$_2$O and Co$_3$O$_4$, although a tiny part of the hydrated phase still retains. The schematic representation of the water cooperated into the structures is illustrated in figure 6.

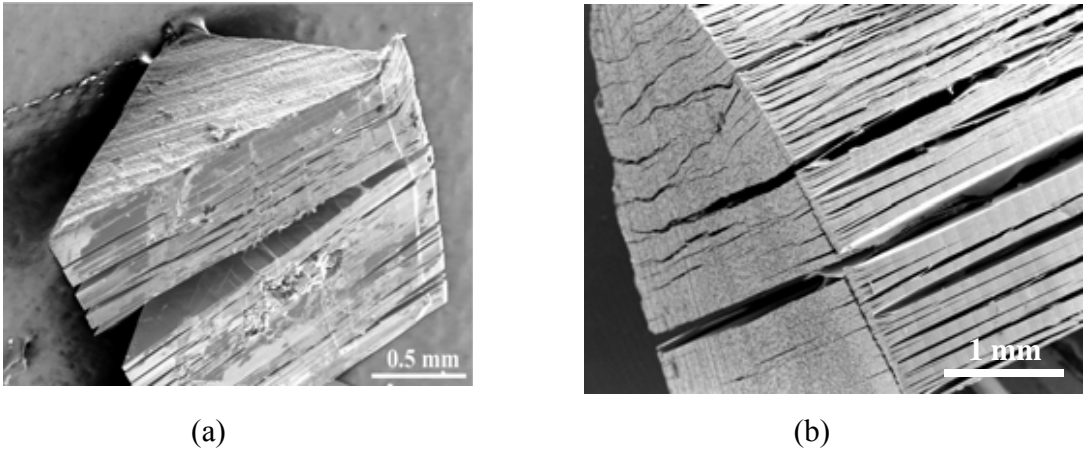

(a)                    (b)

FIG. 4 The typical crystal morphologies showing the layered structure of a $Na_xCoO_2$ single crystal. (a) The cracked layers perpendicular to the $c$ axis of the $Na_{0.3}CoO_2$ crystals after Na-extraction, (b) the "booklet"-like structure of the $Na_{0.3}CoO_2 \cdot 1.3H_2O$ after hydration.

TABLE 1 The possible ordered phases with different water contents **y** in $Na_{0.3}CoO_2 \cdot yH_2O$, their lattice parameters, the corresponding water molecule numbers (***n***) in a formula unit and weight loss ($\Delta w$) during Thermogravimetric experiment (fig. 9 a).

| **y** | 0 | 0.3 | 0.6 | 0.9 | 1.3 | 1.8 |
|---|---|---|---|---|---|---|
| *a* (Å) | 2.827 | 2.820 | 2.823 | 2.825 | 2.823 | 2.818 |
| *c* (Å) | 11.15 | 12.49 | 13.85 | 16.79 | 19.71 | 22.38 |
| ***n*** | 0 | 1 | 2 | 3 | 4 | 6 |
| $\Delta w$(mg) | 28.4 | 23.7 | 19.2 | 13.6 | 9.0 | 0 |

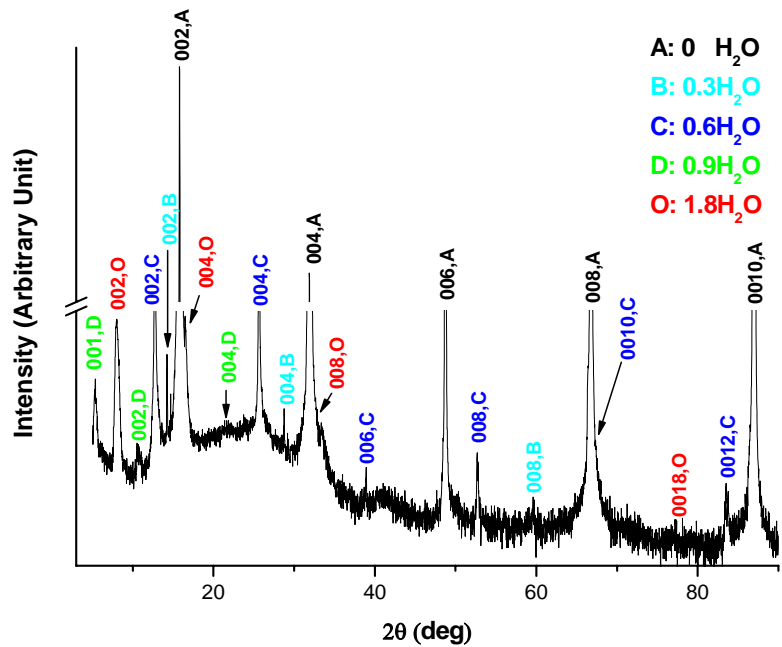

FIG. 5 The *00l* X-ray diffraction patterns showing the mixture of hydrates for Na$_{0.3}$CoO$_2$•yH$_2$O with y=0, 0.3, 0.6, 0.9 and 1.8, respectively.

*3.5 Hydration dynamics*

The high sensitivity of the x-ray diffraction experiment allows us to observe the phase formation as well as its dissolution during the hydration and dehydration of $Na_{0.3}CoO_2$. Figure 7 shows the *002* reflections obtained from a crystal with hydration time from 0 minute up to 10 days. It is not surprising to observe the coexistence of the two phases with y=0 and 0.6 (Fig. 7, 0 min) prior to the hydration, since the non-hydrated sample of $Na_{0.3}CoO_2$ was stored in air and readily absorbs water to form a partial hydrate under ambient conditions. As hydration continues the y=0.6 phase would not vanish until 10 days when a full hydrate phase of y=1.3 is formed. Assuming that the initial diffusion path of the water molecule takes place along the Na-plane to fill the partially occupied sites, this process would not stop until water saturation occurs in these planes. According to the *002* reflection patterns, the expansion of the *c*-axis per intercalant layer is about 1.3 Å averaged from the non-hydrated $Na_{0.3}CoO_2$ (11.2 Å) to the lower hydrate $Na_{0.3}CoO_2 \cdot 0.6H_2O$ (13.8 Å). This value is smaller than the diameter of an oxygen ion (~2.8 Å), suggesting that in this partial hydrate, the Na ions and $H_2O$ molecules accommodate in the same plane. The structural model of y=0.6 is illustrated in figure 6. The evidence in figure 7 indicates that the initial phase formation is $Na_{0.3}CoO_2 \bullet 0.6H_2O$ (y=0.6, n=2). During the different stage of the intercalation process two water molecules were observed to insert in each Na-site and expected to form a cluster of Na $(H_2O)_2$ in the Na layer. A feature in fig. 7 shows that the y=0.6 (n=2) phase reacts like a metastable phase, which can form readily under ambient and humid conditions while only vanishing after a long hydration time of 10 days.

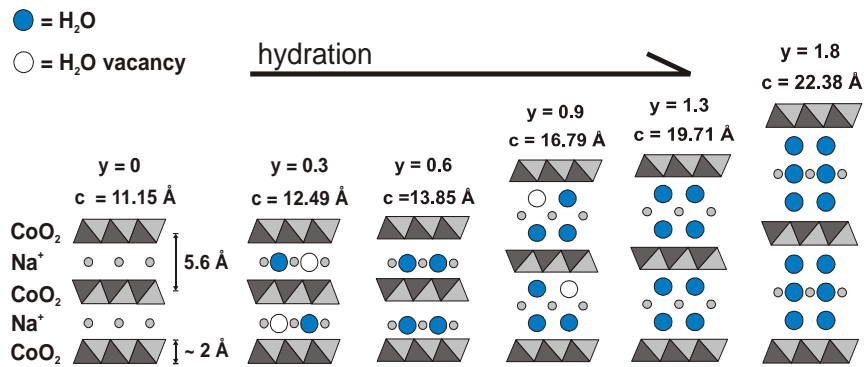

FIG. 6 Schematic representation for the structures of the possible ordered phases of $Na_{0.3}CoO_2 \cdot yH_2O$, y=0, 0.3, 0.6, 0.9, 1.3 and 1.8, which partly referred to [12].

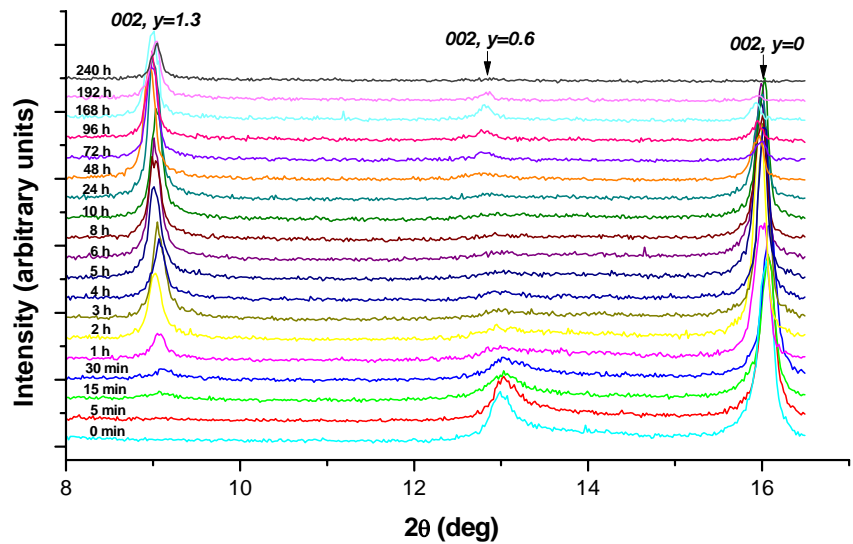

FIG. 7 The *002* reflections show the hydration dynamics of the water molecule intercalation in $Na_{0.3}CoO_2 \bullet yH_2O$. The process indicates two water molecules (y=0.6, n=2) inserted into a formula unit initially and followed by a group of four to form a full hydrate phase (y=1.3, n=4).

With increasing intercalation time the phase of y=1.3 starts to form after a hydration of 15 minutes. The diffraction data in figure 8 show that the *002* intensity of the phase tends to be stronger, indicating more water molecules to accommodate in the lattice and resulting in the increase of phase volume, while vice versa for the non-hydrated phase y=0. During the hydrating process the y=1.3 phase is formed dominantly and completely in 10 days, commencing to a nearly vanishing of both phases of y=0 and 0.6. These results suggest that the sample is fully hydrated with four water molecules, forming clusters of Na $(H_2O)_4$ in the structure. According to the structure illustrated in fig. 3 (b), the phase y=1.3 is formed with four water molecules (two above and two below) to each Na ion and lie in a plane between the Na layer and $CoO_2$ layers. Thus the optimal superconducting phase of $Na_{0.3}CoO_2 \cdot yH_2O$ (y=1.3, n=4) is achieved to have Tc~4.9 K [13]. No evidence of the formation of the phases with y=0.3 (n=1), 0.9 (n=3), or 1.8 (n=6) is observed during the 10-day hydration process. This clearly demonstrates that the initial intercalation process takes place with two water molecules (y=0.6, n=2) and followed by a group of four to form two additional layers between Na and $CoO_2$. A further increase of the hydration time over 15 days may lead to an additional phase of y=1.8 (described in Section 3.6).

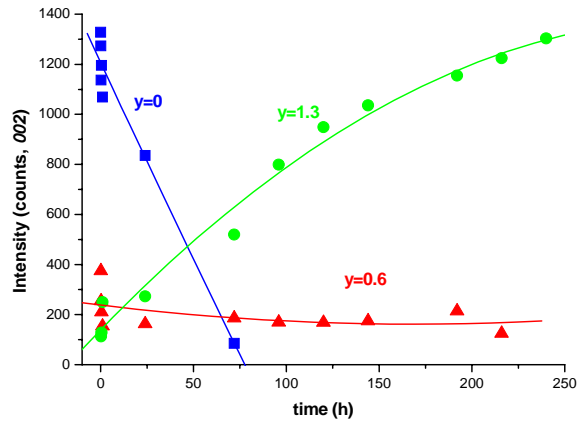

FIG. 8 Time dependence of the *002* counts for y=0, 0.6 and 1.3, respectively, indicating the volume of full hydrate y=1.3 increases while the y=0 phase rapid decreases. The y=0.6 phase is rather stable.

*3.6 Deintercalation process*

After the specimens of $Na_{0.3}CoO_2$ were treated by an over hydration time of 15 days, the dehydration process was then carried out by heating the sample and monitored by the thermogravimetric measurement. The result in figure 9 (a) is the Thermogravimetric curve obtained by the sample heated at 0.3 °C /min in flowing oxygen environment.

There are five plateaus observed in the change of weight via heating sample. Analysis of XRD data suggests that each plateau corresponds to certain dehydrate phases with different water contents, i.e., y values. The y values are calculated from the weight loss and listed in Table 1. Noticeable are turning points between every two plateaus at elevated heating temperatures, which are similar to a phase transition. These can be seen clearly in the derivative of the weight loss curve in the inset of fig. 9 (a). The study of XRD revealed the existence of a majority phase corresponding to each turning point.

The first sharp loss of weight is estimated to be about 13 mg, observed between 28 and 50 °C. This value resulted from the initial water content y~1.8, indicating that the sample is over hydrated with composition $Na_{0.3}CoO_2 \cdot 1.8H_2O$. For an over hydrated sample, it is assumed that six water molecules are filled below, in and above the Na ion site to form a cluster of $Na(H_2O)_6$. The over hydrated structure model is presented with y=1.8, n=6 in figure 6. It was reported [9] that for **y** =1.3 and 1.8 phases the lattice parameters are very close, hence they are difficult to be identified by x-ray diffraction. The XRD data in figure 7 show the reflections of *002, 004* and *008* at 2θ=8.05°, 16.08° and 31.99°, which are resulted from the phase y=1.8 with lattice constant *c*=22.38 Å. This process deintercalated water from **y**=1.8 (n=6) to **y**=0.9 (n=3), corresponding to a loss of about three water molecules from each sodium site. It is notice that the y=1.3 phase is not observable by a turning point in fig. 9. The result is consistent with that of fig. 5. The indication of this rapid loss of water confirms that both phase of over-hydrate (y=1.8, n=6) and optimum hydrate (y=1.3, n=4) are extremely unstable at around room temperature.

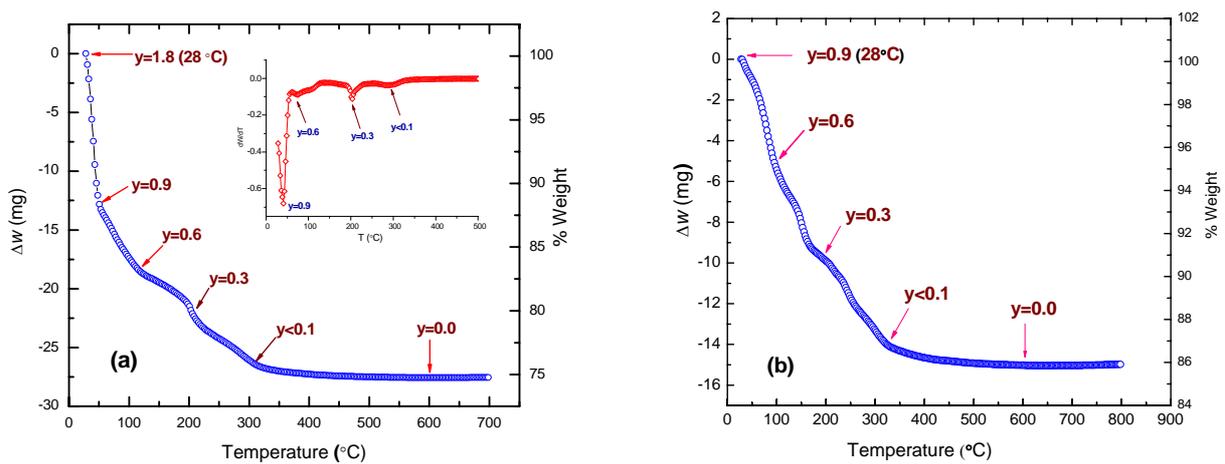

FIG. 9 Thermogravimetric analysis of (a) an over hydrated (15 days) $Na_{0.3}CoO_2 \cdot 1.8H_2O$ (112.9650 mg) showing the temperature dependence of the weight loss for the compound heated at 0.3 °C/min in flowing oxygen. Inset: the derivative curve. (b) The result is reproduced from another sample of partial hydrate $Na_{0.3}CoO_2 \cdot 0.9H_2O$ (107.7800 mg).

We used Raman scattering to further characterize the overhydrated phase. In the used scattering geometry, z(xx)-z, with light polarization within the $CoO_2$ planes, information about two vibrational modes of the $CoO_6$ octahedra is provided. Previous studies on samples with y~1.3 showed an out-of-plane $A_{1g}$ oxygen vibration at about 590 cm$^{-1}$ and an $E_{1g}$ in-plane oxygen vibration at about 480 cm$^{-1}$ [4]. In comparison to these data the two phonon modes are shifted to lower frequency for y=1.8, as shown in figure 10. The present overhydrated sample has been over hydrated and the existence of the y=1.8 phase is confirmed by x-Ray scattering. The observed shift of the phonon frequencies is taken as evidence for a softening of the local binding in the overhydrated phase induced by the increase of the *c* axis parameter and the additional layering (see fig. 6). A similar shift has also been reported comparing the non-hydrated y=0 with the y=1.3 hydrated phase. The phonon frequencies do not show a significant temperature dependence in the temperature range from 10K to RT. There is an additional phonon mode visible at 505 cm$^{-1}$. This mode might be related to another oxygen vibration as the Co-site itself is not Raman-active due to its site inversion symmetry.

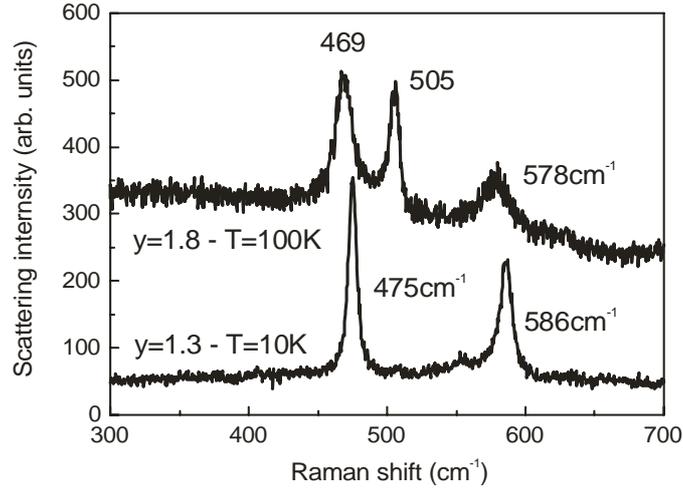

FIG. 10 Raman scattering data on overhydrated $Na_{0.3}CoO_2 \cdot 1.8H_2O$ (upper curve) and $Na_{0.3}CoO_2 \cdot 1.3H_2O$ (lower curve) at low temperatures. The data are shifted for clarity. Peak frequencies of the phonon modes are given.

The following deintercalation processes take place by stages identified as phase transitions from one to another majority phase. At elevated temperatures of approximately 50, 100, 200 and 300 °C, the loss of water is estimated to be one molecule per formula unit, corresponding to the y=0.9 (n=3), y=0.6 (n=2), 0.3 (n=1) and <0.1 (n~0, only trace water), respectively. Their corresponding phase transitions are: $Na_{0.3}CoO_2 \bullet 0.9H_2O$ (50 °C) -> $Na_{0.3}CoO_2 \bullet 0.6H_2O$ (100 °C) -> $Na_{0.3}CoO_2 \bullet 0.3H_2O$ (200 °C) -> $Na_{0.3}CoO_2$ (>300 °C). An entire removal of water from the crystal occurs at about 600 °C. These results are in excellent agreement with a further deintercalation study starting from the partially hydrated phase $Na_{0.3}CoO_2 \bullet 0.9H_2O$. The corresponding data are shown in figure 9 (b). There is an important feature in figures 9 (a) and (b) given by the stepwise decrease of plateau slopes with elevated temperature or progressive deintercalation. This reduction of slope marks a generally higher stability of the compound with lower content of water.

## 4. Conclusion

A frequently addressed problem is the coexistence of both free-water and crystal-water inserted into samples during the process of intercalation. The free-water shows no feature in the XRD measurement, but is observed in the TG curves, which show a sharper loss of the weight in powder samples [3]. Powders consist of a large amount of grain boundaries, giving rise to water inserted into not only crystal lattice but also intergrain spacing during intercalation. As a consequence, both crystal-water and free-water accommodate in the sample and lead to the difficulty in the determination of the **y** values. It is apparent that the results from a single crystal (one grain) provide features with high accuracy. In our TG measurements, the turning points and the plateaus followed are so pronounce that they can be proven by the formation of several majority phases in a transitional state.

Our data show that the water intercalation/deintercalation processes in $Na_{0.3}CoO_2 \bullet yH_2O$ single crystals are essentially different. The intercalation process by immerging the crystals into water takes place in a way of two water molecules inserted initially in Na-plane and followed by a group of four to form a hydrated cluster of Na $(H_2O)_4$ in a formula unit. In the water deintercalation process monitored by TG measurements the water molecules are removed from the hydrated cluster of $Na(H_2O)_4$ in way of one by one at elevated temperature starting from 50 °C. The intercalation/deintercalation dynamics via different hydrates may lead to complex non-equilibrium states in the bulk and especially on the surface of the single crystals. There are indications of multiphase formation and their coexistence in $Na_{0.3}CoO_2 \bullet yH_2O$. Their effect on the electronic and superconducting properties remain unclear. However, depending on the reaction conditions transition-like phenomena mark phase boundaries between quasi-equilibrium states with y=0 (n=0), 0.3 (n=1), 0.6 (n=2), 0.9 (n=3), 1.3 (n=4) and 1.8 (n=6). This signals that, presuming a careful control of hydration conditions, it is possible to study the cobaltates in quasi-equilibrium conditions.

**Acknowledgments**
We thank G. Götz, E. Winckler for technical support.

**Références**
[1] K. Takada, H. Sakurai, E. Takayama-Muromachi, F. Izumi, R. Dilanian, T. Sasaki, Nature, **422,** (2003) 53.
[2] Y. Wang, N.S. Rogado, R.J. Cava, N.P. Ong, Nature **423**, (2003) 425.
[3] R.E. Schaak, T. Klimczuk, M.L. Foo, R.L. Cava, Nature **424**, (2003) 527.
[4] P. Lemmens, V. Gnezdilov, N.N. Kovaleva, K.Y. Choi, H. Sakurai, E. Takayama-Muromachi, K. Takada, T. Sasaki, F.C. Chou, C.T. Lin, B. Keimer, Journ. Phys.: Cond. Mat. **16,** (2004)S857-S865.

[5] J. Cmaidalka, A. Baikalov, Y.Y. Xue, R.L. Meng. And C.W. Chu, Physica C **403**, (2004)125-131.

[6] B. Lorenz, J. Omaidalka, R.L. Meng, and C.W. Chu, Phy. Rev. B **68**, (2003)132504.

[7] H. Sakurai, K.Takada, S. Yoshii, T. Sasaki, K. Kindo, E. Takayama-Muromachi, Phy. Rev. **B 68**, (2003)132507.

[8] S. Bayrakci, C. Bernhard, D. P. Chen, B. Keimer, R. K. Kremer, P. Lemmens, C. T. Lin, C. Niedermayer, J. Strempfer, Phys. Rev. **B 69**, (2004)100410(R).

[9] R. Jin, B.C. Sales. P. Khalifah, D. Mandrus, Phy. Rev. Lett., **91**, (2003) 217001.

[10] J.W. Lynn, Q. Huang, C.M. Brown, V.L. Miller, M.L. Foo, R.E. Schaak, C.Y. Jones, E.A. Mackey, R.J. Cava, Phy. Rev. B **68**, (2003)214516.

[11] J.D. Jorgensen, M. Avdeev, D.G. Hinks, J.C. Burley, S. Short, Phy. Rev. B **68**, (2003) 214517.

[12] M.L. Foo, R.E. Schaak, V.L. Miller, T. Klimczuk, N.S. Rogado, Y. Wang, G.C. Lau, C. Craley, H. W. Zandbergen, N.P. Ong, R.J. Cava, Solid State Commun. **127**, (2003) 33.

[13] D.P. Chen, H.C. Chen, A. Maljuk, A. Kulakov, H. Zhang, P. Lemmens, C.T. Lin, Phys. Rev. B **70**, (2004)024506.

[14] F. Rivadulla, J.-S. Zhou, J. B. Goodenough, Phy. Rev. B **68**, (2003) 075108.

[15] M.D. Johannes, D.J. Singh, Phy. Rev. B **70**, (2004)014507.

[16] M.N. Iliev, A.P. Litvinchuk, R.L. Meng, Y.Y. Sun, J. Cmaidalka, C.W. Chu, Physica C **402**, (2004)239.

[17] F.C. Chou, J.H. Cho, P.A. Lee, E.T. Abel, K.Matan, Y.S. Lee, Phys. Rev. Lett. **92**, (2004)157004.